# Exploring maximally multipartite entanglement of even n qubits with N-tangle

Xin-Wei Zha, Jian-Xia Qi and Yun-Guang Zhang

School of Science, Xi'an University of Posts and Telecommunications, Xi'an 710121,People's Republic of China

**Abstract**   A necessary condition of the maximally multipartite entangled states (MMES) is given via n-tangle. The condition shows that the n-tangle equal zero for the four-, and eight-qubit of MMESs and the n-tangle equal 1 for two- and six- qubits of MMESs . Furthermore，we give the theoretical limitation for maximally ten-qubit and twelve-qubit entangled states. We conjecture that 4n-qubit states of MMES should have n-tangle equal zero and 4n+2-qubit states of MMES should have n-tangle equal one.

Entanglement is considered as the central resource for quantum information and computation [1, 2], and numerous theoretical and experimental works have been done in this field [3–6]. In particular, the search for maximally entangled states has focused a great deal of attention [7–16]. It is then a fundamental question to ask which states are maximally entangled. In the case of 2 qubits, it is known that Bell states are maximally entangled with respect to any measures of entanglement [1].

The concurrence has been shown to be a useful entanglement measure for pure two qubits. In 2000, Coffman, Kundu, and Wootters [7] using concurrence to examine three qubit quantum systems introduced the concept of "residual entanglement", or the 3 – tangle. In 2001, A. Wong and N. Christensen [18] show the 3-tangle was extended to even n qubits, and the extension was called the n-tangle. The n-tangle of even n qubits is invariant under permutations of the qubits, and is an entanglement monotone. In 2010, an expression for four-tangle was obtained by examining the negativity fonts present in a four-way partial transpose under local unitary operations [19].

For higher numbers of qubits, the problem is no longer simple and depends in general on the entanglement measure. In [12], Verstraete et al. refer to maximally entangled states as states with maximally mixed one-qubit reduced density matrices. In [8], Facchi P et al. proposed that the multipartite entanglement of a system of qubits can be characterized in terms of the distribution function of bipartite entanglement (e.g., purity) over all possible bipartitions of the qubits. This led us to formulate the notion of "maximally multipartite entangled states" (MMES), as those states for which average purity (over all balanced bipartitions) is minimal. By their very definition, we introduced a criterion for a maximally multi-qubit entangled state (MMES) with the help of LU transformation invariants. Such a criterion can easily verify several known multi-qubit highly entangled states to be MMES, and in principle can be applied to arbitrary n-qubit entanglement [20]. Also, by the LU transformation we are the first to discover a genuine eight-qubit maximally entangled state proven by the criterion. In this article, we investigate the relation between the entanglement of purity and the tangle of a multipartite system. Then, we study the conditions for obtaining maximally multipartite entangled states (MMES) via N-tangle. Furthermore, we give the theoretical limitation for maximally ten-qubit entangled state.

In 2008, Paolo Facchi et al [8] proposed that the multipartite entanglement of system of qubits can be characterized in terms of distribution function of bipartite over all possible bipartitions of the qubits, namely

$$\pi_{ME} = \binom{n}{n_A}^{-1} \sum_{|A|=n_A} \pi_A, \tag{1}$$

where $n_A = [n/2]$, and Purity reads $\pi_A = Tr_A \rho_A^2$, where $\rho_A = Tr_{\bar{A}} |\psi\rangle\langle\psi|$ is the reduced density matrix of party A. Purity ranges between $\frac{1}{2^{n_A}} \leq \pi_A \leq 1$.

The quantity $\pi_{ME}$ in Eq. (1) measures the average bipartite entanglement over all possible balanced bipartitions. A maximally multipartite entangled state (MMES) is that $\pi_{ME}$ is minimal. We notice that a necessary condition for a state to be a MMES is to be maximally entangled with respect to the N-tangle.

In Ref.[18], Wong and Christensen defined the N-tangle

$$\tau_N = \left|\langle\psi|\sigma_{1y} \otimes \sigma_{2y} \cdots \otimes \sigma_{Ny}|\psi^*\rangle\right|^2. \tag{2}$$

and $\sigma_{iy}$ is the Pauli matrix.

For the n-qubit system, we can express its state vector as

$$|\psi\rangle = \sum a_i |i\rangle. \tag{3}$$

Then, we can obtain[17]

$$\pi_{ME} = C + K(a_i). \tag{4}$$

Where C is a positive constant and $K$ is the function of $a_i$ and can be written as the sum of many square terms, which leads to $K \geq 0$.

Because $K \geq 0$, equation (4) gives a general approach to construct a minimizable structure of $\pi_{ME}$,

by which a state with $K=0$ and therefore $\pi_{ME} = C$ can be proven to be maximally entangled.

For the case of n=2, i.e. the two-qubit system, we can obtain

$$C = \frac{1}{2}, K = \frac{1}{2}(1-\tau_2). \tag{5}$$

where $\tau_2 = |\langle \psi | \sigma_{1y} \otimes \sigma_{2y} | \psi^* \rangle|^2$.

Therefore, if $\tau_2 = 1$, the state is a maximally entangled state.

For n=4 qubits, we have [20] $C = \frac{1}{3}, K = \frac{1}{6}(F_1 + F_2 + F_3 + F_4 + \tau_4),$ (6)

where $F_i = \langle \psi | \hat{\sigma}_{ix} | \psi \rangle^2 + \langle \psi | \hat{\sigma}_{iy} | \psi \rangle^2 + \langle \psi | \hat{\sigma}_{iz} | \psi \rangle^2, \tau_4 = |\langle \psi | \sigma_{1y} \otimes \sigma_{2y} \otimes \sigma_{3y} \otimes \sigma_{4y} | \psi^* \rangle|^2$.

For product state, $F_1 = F_2 = F_3 = F_4 = 1, \tau_4 = 0$, $K = \frac{2}{3}$; for GHZ state, $F_1 = F_2 = F_3 = F_4 = 0, \tau_4 = 1$, $K = \frac{1}{6}$. It is obvious that, for maximally n=4 qubits, it must be $F_1 = F_2 = F_3 = F_4 = 0, \tau_4 = 0$. Therefore, $\tau_4 = 0$ is the necessary condition of maximally n=4 qubits.

For n=6 qubits, we have

$$C = \frac{1}{8}, K = \frac{3}{40}(F_1 + \cdots + F_6) + \frac{1}{40}(F_{12} + \cdots + F_{56}) + \frac{1}{20}(1-\tau_6). \tag{7}$$

Where

$$\begin{aligned} F_{ij} &= \langle \psi | \hat{\sigma}_{ix} \hat{\sigma}_{jx} | \psi \rangle^2 + \langle \psi | \hat{\sigma}_{ix} \hat{\sigma}_{jy} | \psi \rangle^2 + \langle \psi | \hat{\sigma}_{ix} \hat{\sigma}_{jz} | \psi \rangle^2 \\ &+ \langle \psi | \hat{\sigma}_{iy} \hat{\sigma}_{jx} | \psi \rangle^2 + \langle \psi | \hat{\sigma}_{iy} \hat{\sigma}_{jy} | \psi \rangle^2 + \langle \psi | \hat{\sigma}_{iy} \hat{\sigma}_{jz} | \psi \rangle^2 \\ &+ \langle \psi | \hat{\sigma}_{iz} \hat{\sigma}_{jx} | \psi \rangle^2 + \langle \psi | \hat{\sigma}_{iz} \hat{\sigma}_{jy} | \psi \rangle^2 + \langle \psi | \hat{\sigma}_{iz} \hat{\sigma}_{jz} | \psi \rangle^2 \end{aligned}$$

$$\tau_6 = |\langle \psi | \sigma_{1y} \otimes \sigma_{2y} \otimes \sigma_{3y} \otimes \sigma_{4y} \otimes \sigma_{5y} \otimes \sigma_{6y} | \psi^* \rangle|^2 \tag{8}$$

For product state, $F_1 = F_2 \cdots = F_6 = 1, F_{12} = \cdots = F_{56} = 1, \tau_6 = 0, K = \frac{7}{8}$;

for GHZ state, $F_1 = F_2 \cdots = F_6 = 0, F_{12} = \cdots = F_{56} = 1, \tau_6 = 1, K = \frac{3}{8}$;

For maximally n=6 qubits, it must be $F_1 = F_2 \cdots = F_6 = 0$, $F_{12} = \cdots = F_{56} = 0$, $\tau_6 = 1$.

Therefore, $\tau_6 = 1$ is the necessary condition of maximally n=6 qubits.

For n=8 qubits, we have $C = \frac{6}{70}$,

$$K = \frac{11}{280}(F_1 + \cdots + F_8) + \frac{1}{70}(F_{12} + \cdots + F_{78}) + \frac{1}{280}(F_{123} + \cdots + F_{678}) + \frac{1}{70}\tau_8 \quad (9)$$

where

$$F_{ijk} = \langle\psi|\hat{\sigma}_{ix}\hat{\sigma}_{jx}\hat{\sigma}_{kx}|\psi\rangle^2 + \langle\psi|\hat{\sigma}_{ix}\hat{\sigma}_{jx}\hat{\sigma}_{ky}|\psi\rangle^2 + \langle\psi|\hat{\sigma}_{ix}\hat{\sigma}_{jx}\hat{\sigma}_{kz}|\psi\rangle^2$$
$$+ \langle\psi|\hat{\sigma}_{ix}\hat{\sigma}_{jy}\hat{\sigma}_{kx}|\psi\rangle^2 + \langle\psi|\hat{\sigma}_{ix}\hat{\sigma}_{jy}\hat{\sigma}_{ky}|\psi\rangle^2 + \langle\psi|\hat{\sigma}_{ix}\hat{\sigma}_{jy}\hat{\sigma}_{kz}|\psi\rangle^2$$
$$+ \cdots$$
$$+ \langle\psi|\hat{\sigma}_{iz}\hat{\sigma}_{jz}\hat{\sigma}_{kx}|\psi\rangle^2 + \langle\psi|\hat{\sigma}_{iz}\hat{\sigma}_{jz}\hat{\sigma}_{ky}|\psi\rangle^2 + \langle\psi|\hat{\sigma}_{iz}\hat{\sigma}_{jz}\hat{\sigma}_{kz}|\psi\rangle^2$$

$$\tau_8 = |\langle\psi|\sigma_{1y} \otimes \sigma_{2y} \otimes \sigma_{3y} \otimes \sigma_{4y} \otimes \sigma_{5y} \otimes \sigma_{6y} \otimes \sigma_{7y} \otimes \sigma_{8y}|\psi^*\rangle|^2.$$

For product state, $F_1 = F_2 \cdots = F_8 = 1$, $F_{12} = \cdots = F_{78} = 1$, $F_{123} = \cdots = F_{678} = 1$, $\tau_8 = 0$, $K = \frac{64}{70}$; for GHZ state, $F_1 = F_2 \cdots = F_8 = 0$, $F_{12} = \cdots = F_{78} = 1$, $F_{123} = \cdots = F_{678} = 0$, $\tau_8 = 1$, $K = \frac{29}{70}$. For maximally n=8 qubits, it must be $F_1 = F_2 \cdots = F_8 = 0$, $F_{12} = \cdots = F_{78} = 0$, $F_{123} = \cdots = F_{678} = 0$, $\tau_8 = 0$. The state searched for in [20],

$$|\psi_M\rangle_{12345678} =$$
$$\frac{1}{8}\{[(|000\rangle+|111\rangle)|0\rangle+(|010\rangle+|101\rangle)|1\rangle][(|000\rangle-|111\rangle)|0\rangle+(|000\rangle-|111\rangle)|1\rangle]$$
$$+[(|001\rangle+|110\rangle)|0\rangle+(|000\rangle-|111\rangle)|1\rangle][(|000\rangle+|111\rangle)|0\rangle+(|100\rangle+|011\rangle)|1\rangle]$$
$$+[(|010\rangle+|101\rangle)|0\rangle+(|100\rangle+|011\rangle)|1\rangle][(|000\rangle+|111\rangle)|0\rangle-(|010\rangle+|101\rangle)|1\rangle]$$
$$+[(|000\rangle-|111\rangle)|0\rangle+(|010\rangle-|101\rangle)|1\rangle][(|000\rangle-|111\rangle)|0\rangle+(|100\rangle-|101\rangle)|1\rangle]\}$$

It can be showed that $F_1 = F_2 \cdots = F_8 = 0$,

$$F_{12} = \cdots = F_{78} = 0, \quad F_{123} = \cdots = F_{678} = 0, \quad \tau_8 = 0.$$

Therefore, $\tau_8 = 0$ is the necessary condition of maximally n=8 qubits.

For n=10 qubits, we have

$$C = \frac{13}{336},$$

$$K = \frac{5}{252}(F_1 + \cdots + F_{10}) + \frac{2}{252}(F_{12} + \cdots + F_{9,10}) + \frac{1}{252} \times \frac{5}{8}(F_{123} + \cdots + F_{89,10})$$
$$+ \frac{2}{252} \times \frac{1}{8}(F_{1234} + \cdots + F_{7,8,9,10}) + \frac{1}{252}(1 - \tau_{10}) \quad (10)$$

where

$$F_{ijkl} = \langle\psi|\hat{\sigma}_{ix}\hat{\sigma}_{jx}\hat{\sigma}_{kx}\hat{\sigma}_{lx}|\psi\rangle^2 + \langle\psi|\hat{\sigma}_{ix}\hat{\sigma}_{jx}\hat{\sigma}_{kx}\hat{\sigma}_{ly}|\psi\rangle^2 + \langle\psi|\hat{\sigma}_{ix}\hat{\sigma}_{jx}\hat{\sigma}_{kx}\hat{\sigma}_{lz}|\psi\rangle^2$$
$$+ \langle\psi|\hat{\sigma}_{ix}\hat{\sigma}_{jx}\hat{\sigma}_{ky}\hat{\sigma}_{lx}|\psi\rangle^2 + \langle\psi|\hat{\sigma}_{ix}\hat{\sigma}_{jx}\hat{\sigma}_{ky}\hat{\sigma}_{ly}|\psi\rangle^2 + \langle\psi|\hat{\sigma}_{ix}\hat{\sigma}_{jx}\hat{\sigma}_{ky}\hat{\sigma}_{lz}|\psi\rangle^2 \quad (11)$$
$$+ \cdots$$
$$+ \langle\psi|\hat{\sigma}_{iz}\hat{\sigma}_{jz}\hat{\sigma}_{kz}\hat{\sigma}_{lx}|\psi\rangle^2 + \langle\psi|\hat{\sigma}_{iz}\hat{\sigma}_{jz}\hat{\sigma}_{kz}\hat{\sigma}_{ly}|\psi\rangle^2 + \langle\psi|\hat{\sigma}_{iz}\hat{\sigma}_{jz}\hat{\sigma}_{kz}\hat{\sigma}_{lz}|\psi\rangle^2$$

and

$$\tau_{10} = \left|\langle\psi|\sigma_{1y} \otimes \sigma_{2y} \otimes \sigma_{3y} \otimes \sigma_{4y} \otimes \sigma_{5y} \otimes \sigma_{6y} \otimes \sigma_{7y} \otimes \sigma_{8y} \otimes \sigma_{9y} \otimes \sigma_{10y}|\psi^*\rangle\right|^2.$$

For product state, $F_1 = F_2 \cdots = F_{10} = 1$, $F_{12} = \cdots = F_{9,10} = 1$, $F_{123} = \cdots = F_{8,9,10} = 1$, $F_{1234} = \cdots = F_{7,8,9,10} = 1$, $\tau_{10} = 0$, $K = \frac{323}{336}$; for GHZ state, $F_1 = F_2 \cdots = F_8 = 0$, $F_{12} = \cdots = F_{9,10} = 1$, $F_{123} = \cdots = F_{678} = 0$, $F_{1234} = \cdots = F_{7,8,9,10} = 1$, $\tau_{10} = 1$, $K = \frac{155}{336}$.

Hence, for n=1 qubits, if $F_1 = F_2 \cdots = F_8 = 0$, $F_{12} = \cdots = F_{78} = 0$, $F_{123} = \cdots = F_{678} = 0$, $\tau_{10} = 1$. Then $K = 0$. Then $\pi_{ME} = \frac{13}{336}$

That is to say, for maximally n=10 qubits, the theoretical limitation for which over all balanced bipartitions is minimal is $\pi_{ME} = \frac{13}{336}$. We can find that $\frac{1}{32} \prec \frac{13}{336} \prec \frac{1}{16}$.

For n=12 qubits, we have

$$C = \frac{157}{7392},$$

$$K = \frac{37}{3696}(F_1 + \cdots + F_{12}) + \frac{31}{7392}(F_{12} + \cdots + F_{11,12}) + \frac{11}{7392}(F_{123} + \cdots + F_{10,11,12})$$
$$+ \frac{3}{7392}(F_{1234} + \cdots + F_{9,10,11,12}) + \frac{1}{7392} \times \frac{1}{2}(F_{1234} + \cdots + F_{9,10,11,12}) + \frac{\tau_{12}}{924} \quad (12)$$

and $\tau_{12} = \left|\langle\psi|\sigma_{1y} \otimes \sigma_{2y} \otimes \cdots \otimes \sigma_{11y} \otimes \sigma_{12y}|\psi^*\rangle\right|^2.$

For product state, $F_1 = F_2 \cdots = F_{12} = 1$, $F_{12} = \cdots = F_{11,12} = 1$, $F_{123} = \cdots = F_{10,11,12} = 1$, $F_{1234} = \cdots = F_{9,10,11,12} = 1$, $F_{12345} = \cdots = F_{8,9,10,11,12} = 1$, $\tau_{12} = 0$, $K = \frac{323}{336}$; for GHZ state, $F_1 = F_2 \cdots = F_8 = 0$, $F_{12} = \cdots = F_{78} = 1$, $F_{123} = \cdots = F_{678} = 0$, $F_{1234} = \cdots = F_{9,10,11,12} = 1$, $F_{12345} = \cdots = F_{8,9,10,11,12} = 0$, $\tau_{12} = 1$, $K = \frac{155}{336}$.

Hence, for n=12 qubits,

if $F_1 = F_2 \cdots = F_8 = 0$, $F_{12} = \cdots = F_{11,12} = 0$, $F_{1234} = \cdots = F_{9,10,11,12} = 0$, $F_{12345} = \cdots = F_{8,9,10,11,12} = 0$, $\tau_{12} = 0$. Then $K = 0$. Therefore $\pi_{ME} = \frac{157}{7392}$.

That is to say, for maximally n=12 qubits, the theoretical limitation for which over all balanced bipartitions is minimal is $\pi_{ME} = \frac{157}{7392}$. We can find that $\frac{1}{64} < \frac{13}{336} < \frac{1}{32}$.

In summary, we first introduced a relationship between the n-tangle and the multipartite entanglement of even n qubits, furthermore, a criterion for a maximally multi-qubit of even n qubits by n-tangle and LU transformation invariants is given. Such a criterion can easily verify several known multi-qubit of even n qubits states to be MMES, and in principle can be applied to arbitrary even n-qubit entanglement. We have revealed that the n-tangle equal zero for four and eight qubits of MMESs and for two, six and ten qubits of MMESs, the n-tangle equal 1 is the necessary condition. We believe this criterion can play an important role in determining whether a state discovered in future is maximally entangled, We conjecture that 4n qubit states of maximum entanglement should have n-tangle equal zero and 4n+2 qubit states of maximum entanglement should have n-tangle equal one.

Acknowledgements

This work is supported by the Natural Science Foundation of Shaanxi Province of China (Grant No. 2013JM1009).